**After DART: Using the first full-scale test of a kinetic impactor to inform a future planetary defense mission**

Thomas S. Statler[1], Sabina D. Raducan[2], Olivier S. Barnouin[3], Mallory E. DeCoster[3], Steven R. Chesley[4], Brent Barbee[5], Harrison F. Agrusa[6], Saverio Cambioni[7], Andrew F. Cheng[3], Elisabetta Dotto[8], Siegfried Eggl[9], Eugene G. Fahnestock[4], Fabio Ferrari[2], Dawn Graninger[3], Alain Herique[10], Isabel Herreros[11], Masatoshi Hirabayashi[12,13], Stavro Ivanovski[14], Martin Jutzi[2], Özgür Karatekin[15], Alice Lucchetti[16], Robert Luther[17], Rahil Makadia[9], Francesco Marzari[18], Patrick Michel[19], Naomi Murdoch[20], Ryota Nakano[13], Jens Ormö[11], Maurizio Pajola[16], Andrew S. Rivkin[3], Alessandro Rossi[21], Paul Sánchez[22], Stephen R. Schwartz[23], Stefania Soldini[24], Damya Souami[19], Angela Stickle[3], Paolo Tortora[25], Josep M. Trigo-Rodríguez[26,27], Flaviane Venditti[28], Jean-Baptiste Vincent[29], and Kai Wünnemann[17,30]

[1] Planetary Defense Coordination Office and Planetary Science Division, NASA Headquarters, 300 Hidden Figures Way SW, Washington DC 20546, USA; Thomas.S.Statler@nasa.gov
[2] Space Research and Planetary Sciences, Physics Institute, University of Bern, Bern, 3012, Switzerland
[3] Johns Hopkins University Applied Physics Laboratory, Laurel MD 20723, USA
[4] Jet Propulsion Laboratory, California Institute of Technology, Pasadena CA 91109, USA
[5] NASA Goddard Space Flight Center, Greenbelt MD 20771, USA
[6] Department of Astronomy, University of Maryland, College Park, MD 20742, USA
[7] Department of Earth, Atmospheric & Planetary Sciences, Massachusetts Institute of Technology, Cambridge, MA, USA
[8] INAF-Osservatorio Astronomico di Roma, Rome, I-00078, Italy
[9] Department of Aerospace Engineering, University of Illinois at Urbana-Champaign, Urbana, IL 61801, USA
[10] Univ. Grenoble Alpes, CNRS, CNES, IPAG, 38000 Grenoble, France
[11] Centro de Astrobiología CSIC-INTA, Instituto Nacional de Técnica Aeroespacial, 28850 Torrejón de Ardoz, Spain
[12] Department of Geosciences, Auburn University, Auburn AL 36849, USA
[13] Department of Aerospace Engineering, Auburn University, Auburn AL 36849, USA
[14] INAF- Osservatorio Astronomico di Trieste, Trieste 34143, Italy
[15] Royal Observatory of Belgium, Belgium
[16] INAF-Astronomical Observatory of Padova, Padova 35122, Italy
[17] Museum für Naturkunde – Leibniz Institute for Evolution and Biodiversity Science
[18] University of Padova, Padova, Italy
[19] Université Côte d'Azur, Observatoire de la Côte d'Azur, CNRS, Laboratoire Lagrange, Nice 06304, France
[20] Institut Supérieur de l'Aéronautique et de l'Espace (ISAE-SUPAERO), Université de Toulouse, Toulouse, France
[21] IFAC-CNR, Sesto Fiorentino 50019, Italy
[22] Colorado Center for Astrodynamics Research, University of Colorado Boulder, Boulder CO 80303, USA
[23] Planetary Science Institute, Tucson, AZ 85719, USA
[24] Department of Mechanical, Materials and Aerospace Engineering, University of Liverpool, Liverpool, United Kingdom
[25] Alma Mater Studiorum – Università di Bologna, Department of Industrial Engineering, Interdepartmental Center for Industrial Research in Aerospace, Via Fontanelle 40 – Forlì (FC) – I-47121, Italy
[26] Institute of Space Sciences (ICE, CSIC), Cerdanyola del Vallès, 08193 Barcelona, Catalonia, Spain
[27] Institut d'Estudis Espacials de Catalunya (IEEC), Ed. Nexus, 08034 Barcelona, Catalonia, Spain
[28] Arecibo Observatory, University of Central Florida, HC-3 Box 53995, Arecibo, PR 00612, USA
[29] German Aerospace Center, DLR Berlin, Germany
[30] Freie Universität Berlin, Germany

**ABSTRACT**

 NASA's Double Asteroid Redirection Test (DART) is the first full-scale test of an asteroid deflection technology. Results from the hypervelocity kinetic impact and Earth-based observations, coupled with LICIACube and the later Hera mission, will result in measurement of the momentum transfer efficiency accurate to ~10% and characterization of the Didymos binary system. But DART is a single experiment; how could these results be used in a future planetary defense necessity involving a different asteroid? We examine what aspects of Dimorphos's response to kinetic impact will be constrained by DART results; how these constraints will help refine knowledge of the physical properties of asteroidal materials and predictive power of impact simulations; what information about a potential Earth impactor could be acquired before a deflection effort; and how design of a deflection mission should be informed by this understanding. We generalize the momentum enhancement factor $\beta$, showing that a particular direction-specific $\beta$ will be directly determined by the DART results, and that a related direction-specific $\beta$ is a figure of merit for a kinetic impact mission. The DART $\beta$ determination constrains the ejecta momentum vector, which, with hydrodynamic simulations, constrains the physical properties of Dimorphos's near-surface. In a hypothetical planetary defense exigency, extrapolating these constraints to a newly discovered asteroid will require Earth-based observations and benefit from in-situ reconnaissance. We show representative predictions for momentum transfer based on different levels of reconnaissance and discuss strategic targeting to optimize the deflection and reduce the risk of a counterproductive deflection in the wrong direction.

# 1    INTRODUCTION

NASA's Double Asteroid Redirection Test (DART) mission is humanity's first attempt to alter the path of a natural celestial body in space. At the time of this writing, the DART spacecraft is en route to the binary asteroid system (65803) Didymos, where it will collide with the secondary body Dimorphos on 2022 September 26, at a speed of approximately 6 km/s.

DART's objective (Cheng et al. 2016, Cheng et al., 2018, Rivkin et al., 2021) is to demonstrate and assess the technical efficacy of hypervelocity kinetic impact as an asteroid deflection technology for defending Earth against the natural hazard posed by near-Earth objects (NEOs). The word "double" in the mission's name has a double meaning: in addition to encountering a double (binary) asteroid, DART is conducting a double test. The first test assesses the design of the spacecraft and its systems, the execution of autonomous terminal navigation, and whether a kinetic impact can be reliably and precisely achieved. The second test assesses the response of the asteroid to the kinetic impact and the efficiency with which it is deflected. The first test ends, and the second begins, at the moment of impact, when the flight hardware is deliberately obliterated.

The details of the DART mission, the known properties of Didymos and Dimorphos, and expected outcomes from the double test are presented in several current publications (e.g., Rivkin et al. 2021, Richardson et al. 2022, Stickle et al. 2022, Daly et al. 2022). The spacecraft carries a single instrument, the Didymos Reconnaissance & Asteroid Camera for OpNav (DRACO; Fletcher et al. 2018), which acquires images for the autonomous navigation system (the Small-body Maneuvering Autonomous Real-Time Navigation, or "SMARTNav") and for the Investigation Team's analysis. During the terminal approach phase DRACO acquires and returns images at a rate of approximately one per second. The highest-resolution images are required to have a pixel scale of 50 cm/pixel or finer, to be achieved roughly 16 seconds prior to DART's impact. Although it is not known when the final image returned to Earth will be taken, it is thought that an image at 9 cm/pixel may be successfully returned.

Piggy-backing on the DART spacecraft is the Italian Space Agency (ASI) Light Italian Cubesat for Imaging of Asteroids (LICIACube; Dotto et al. 2021). Released 15 days before impact, LICIACube observes the binary asteroid and the DART impact from a close approach (CA) distance of ~ 51 km, approximately 3 minutes behind DART, with two optical imagers: the LICIACube Explorer Imaging for Asteroid (LEIA), a high-resolution panchromatic camera; and the LICIACube Unit Key Explorer (LUKE), a wide-angle 3-band color imager. LEIA achieves a spatial scale < 2 m/pixel at close approach, complementing DRACO. LICIACube's objectives are to verify the DART impact and image the impact site, to image the ejecta plume from multiple angles, and to image the non-impact hemisphere of Dimorphos.

The DART investigation exists within the broader NASA- and ESA-supported international collaboration known as the Asteroid Impact Deflection Assessment (AIDA), and the mission is strongly synergistic with ESA's Hera mission (Michel et al. 2022), which is planned to arrive at Didymos in 2026. Hera's remote sensing instruments and two deployable cubesats will observe both asteroids for at least 6 months, obtaining global coverage of each object to a resolution of 10 cm/pixel, including measurements of the size and shape of the crater left by DART. Radio tracking of Hera from Earth and the satellite-to-satellite range-rate tracking between Hera and the deployed Juventas cubesat will enable the masses of Didymos and Dimorphos to be determined to accuracies of 0.00001% and 0.1%, respectively. Juventas and its sibling cubesat Milani, both of which are equipped with accelerometers, are intended to eventually land on Dimorphos, enabling direct (albeit momentary) measurements of surface strength.

The essence of kinetic impact is not merely that the impactor spacecraft deposits its own momentum on the target, but that the kinetic energy liberated in the hypervelocity collision excavates and ejects a large amount of material, the recoil impulse from which can exceed the directly delivered momentum, greatly enhancing the effect of the deflection. A central result from DART, refined using later results from Hera, will be the determination of the total momentum imparted to the target asteroid, and with it, a measure of the momentum transfer efficiency. This is a determination that cannot be made on the ground because of the unavailability of large amounts of representative asteroidal materials and the impossibility of sustained microgravity conditions in the laboratory.

The momentum transfer efficiency is also the key quantity that would govern the effectiveness of a future kinetic impactor, should one ever have to be designed and deployed in response to an imminent Earth-impact danger. The planetary defense community ought to have the capacity to predictively constrain this efficiency using information about the threatening object obtainable after its discovery. And yet, DART is only a single experiment on a single asteroid that could turn out to be quite different from the actual object needing deflection.

In this paper we explicitly focus on this hypothetical future situation, and consider how the results from the DART experiment may be used, both for improving the understanding of the response of asteroids to hypervelocity impacts, and for informing a possible future situation where a kinetic impact deflection may be necessary. Extrapolating from the single case of Dimorphos to a still-undiscovered object of undetermined nature is extremely challenging, and we are forced to treat the problem in general terms. We start by generalizing the long-used definition of the momentum transfer efficiency $\beta$, and show how a $\beta$ factor can be defined associated with any arbitrary direction in space. We show that $\beta_p$, the efficiency associated with the orbital motion of Dimorphos at the moment of impact, follows directly from the DART measurements and constrains the ejecta response of the surface; and that $\beta_u$, the efficiency associated with the optimal deflection direction for a potential Earth-impacting object, can serve as a figure of merit for the design of a future kinetic impact deflection attempt. Next, we summarize the current state of knowledge, from laboratory experiment and numerical simulation, regarding the expected ejecta response and its relationship to the physical properties of the asteroid surface. We also discuss what will be needed to develop the computational capability to accurately predict the impact cratering process and thus the ejecta response for a given asteroid surface and impactor parameters. Then, we consider the problem of characterizing a newly discovered, dangerous asteroid, and what could be learned from different levels of advance reconnaissance. With this in mind, we give representative examples of the types of quantitative predictions that could be made for kinetic impact outcomes, and consider how strategic targeting could optimize the deflection and/or mitigate the risk of a counterproductive result that pushes the asteroid in the wrong direction. Finally, we discuss the further considerations that would be needed in the case of a rapidly rotating or binary asteroid threat.

## 2  MOMENTUM TRANSFER EFFICIENCIES

### 2.1  A Bevy of Betas

In the standard 1-dimensional "toy model" of kinetic impact, an incoming spacecraft impacts a spherical asteroid dead-center, giving it a push in the spacecraft's direction of motion, enhanced by the recoil from ejecta whose total momentum is directed back along the spacecraft's path. The momentum enhancement factor $\beta$ is defined (Holsapple 2004) as the constant of proportionality between the incoming spacecraft's momentum at infinity, $p_s$, and the outgoing final change in momentum of the asteroid, $\Delta p$:

$$\Delta p = \beta p_s. \tag{1}$$

It is fully acknowledged in the literature that the true situation is 3-dimensional (Cheng et al. 2020, Rivkin et al. 2021), but it is still conventional (e.g., Holsapple & Housen 2012, Feldhacker et al. 2017) to adopt a 1-dimensional definition involving the normal components of the momenta, $\Delta p_n = \hat{\boldsymbol{n}} \cdot \Delta \boldsymbol{p}$ and $p_{sn} = \hat{\boldsymbol{n}} \cdot \boldsymbol{p}_s$ (where $\hat{\boldsymbol{n}}$ is the surface normal unit vector at the point of impact, and $\boldsymbol{p}_s$ and $\Delta \boldsymbol{p}$ are the vector spacecraft momentum[31] and asteroid momentum change, respectively):

$$\Delta p_n = \beta_n p_{sn}. \tag{2}$$

The normal-component enhancement factor $\beta_n$ is what is typically denoted $\beta$ in the literature, and the DART project adopts this definition for consistency with standard practice in the field (Cheng, et al. 2016, Rivkin et al. 2021). In this paper we will keep the subscript $n$ on $\beta_n$ for clarity.

Looking beyond DART, however, there is value in retaining the full three-dimensionality of momentum transfer. In analogy with equation (1), we define $\mathbf{B}$ (capital beta) as the tensor that maps the spacecraft momentum vector $\boldsymbol{p}_s$ into the asteroid momentum change vector $\Delta \boldsymbol{p}$:

$$\Delta \boldsymbol{p} = \mathbf{B} \cdot \boldsymbol{p}_s. \tag{3}$$

This relation is fully general and assumes no particular geometry. The correspondence with the normal-component definition can be seen by taking the scalar product of equation (3) with $\hat{\boldsymbol{n}}$:

$$\hat{\boldsymbol{n}} \cdot \Delta \boldsymbol{p} = \hat{\boldsymbol{n}} \cdot \mathbf{B} \cdot \boldsymbol{p}_s. \tag{4}$$

---

[31] In this paper we ignore the distinction between the spacecraft momentum at infinity and at impact, because in any viable kinetic impact scenario, the spacecraft speed vastly exceeds the escape speed from the target body, so the difference will be negligible.

Then, by comparison with equation (2),

$$\beta_n = \frac{\hat{\boldsymbol{n}} \cdot \mathbf{B} \cdot \hat{\boldsymbol{p}}_s}{\hat{\boldsymbol{n}} \cdot \hat{\boldsymbol{p}}_s}, \tag{5}$$

where $\hat{\boldsymbol{p}}_s$ is the unit vector representing the spacecraft's incoming direction. But equation (5) represents only one out of an arbitrary number of different possible "betas", since for any direction $\hat{\boldsymbol{u}}$, one can define an associated $\beta_u$:

$$\beta_u = \frac{\hat{\boldsymbol{u}} \cdot \mathbf{B} \cdot \hat{\boldsymbol{p}}_s}{\hat{\boldsymbol{u}} \cdot \hat{\boldsymbol{p}}_s}. \tag{6}$$

To see the value of such a quantity, consider the fundamental measurement made in the DART experiment: the change in the period $P$ of the binary orbit. In the approximation where the orbit is Keplerian, this period change corresponds to a change in semimajor axis $a$, and therefore to a change in orbital energy per unit mass $\mathcal{E}$ given by

$$\Delta \mathcal{E} = -\frac{2}{3} \frac{\mathcal{E}}{P} \Delta P = \left( \frac{4\pi^2 G^2 M^2}{27 \, P^5} \right)^{1/3} \Delta P, \tag{7}$$

where the far-right-hand side follows from Kepler's laws and applies to the case of a circular pre-impact orbit.[32] The quantity in parentheses (in which $M$ is the total system mass and $G$ is the gravitational constant) is currently known to approximately 7.5% accuracy (Richardson et al. 2022) from ground-based observations. Assuming that the momentum transfer is quick enough that it can be taken to occur at constant binary separation and therefore constant potential energy, and that the fractional changes are all small, as expected, then, using equations (3) and (6),

$$\Delta \mathcal{E} = \frac{\boldsymbol{p} \cdot \Delta \boldsymbol{p}}{m^2} = \frac{p p_s}{m^2} \hat{\boldsymbol{p}} \cdot \mathbf{B} \cdot \hat{\boldsymbol{p}}_s = \frac{m_s}{m} (\boldsymbol{v} \cdot \boldsymbol{v}_s) \beta_p, \tag{8}$$

where $m_s$ is the spacecraft mass, $m$ and $\boldsymbol{p}$ are the secondary's mass and momentum in the frame of the binary barycenter, $\hat{\boldsymbol{p}}$ is the unit vector in the direction of $\boldsymbol{p}$, $\boldsymbol{v}$ and $\boldsymbol{v}_s$ are the secondary and spacecraft velocities in that frame, and $\beta_p$ is the momentum enhancement in the direction of the secondary's motion at the moment of impact. For the DART experiment, the spacecraft mass and velocity and Dimorphos's velocity are known, and Dimorphos's mass must be estimated – but *will* be known accurately following the Hera mission's full characterization of the Didymos system. Thus $\beta_p$ is directly calculable from the parameters of the impact and the measured period change, using equations (7) and (8):

$$\beta_p = \frac{m}{m_s} \left( \frac{4\pi^2 G^2 M^2}{27 \, P^5} \right)^{1/3} \frac{\Delta P}{\boldsymbol{v} \cdot \boldsymbol{v}_s}. \tag{9}$$

Note that it is $\beta_p$, not $\beta_n$, that is *directly* constrained by DART observations. For comparison, a closed-form equation for $\beta_n$ is given in equation (B-3) of Rivkin, et al. (2021). With modifications for notational consistency with this paper, it is:

$$\beta_n = \frac{\frac{m}{m_s} \Delta V_T - \boldsymbol{V}_{\infty \perp n} \cdot \hat{\boldsymbol{p}} + V_{\infty n} \boldsymbol{\epsilon} \cdot \hat{\boldsymbol{p}}}{V_{\infty n} (\hat{\boldsymbol{n}} + \boldsymbol{\epsilon}) \cdot \hat{\boldsymbol{p}}}, \tag{10}$$

where $\Delta V_T$ is the component of Dimorphos's velocity change in the direction of its orbital motion, $V_{\infty n} \equiv \hat{\boldsymbol{n}} \cdot \boldsymbol{v}_s$, and $\boldsymbol{V}_{\infty \perp n} \equiv \boldsymbol{v}_s - V_{\infty n} \hat{\boldsymbol{n}}$. The vector $\boldsymbol{\epsilon}$, parametrizing the offset of the ejecta momentum from the surface normal, is perpendicular to the normal $\hat{\boldsymbol{n}}$ and has magnitude $\tan \alpha'$, where $\alpha'$ is the angle between $\hat{\boldsymbol{n}}$ and the direction of the ejecta momentum. Equation (10), even though exact, is more complicated and more challenging in application than equation (9), in part because $\beta_n$ relies on the direction of the ejecta momentum, which is not directly measured. As a result, $\beta_n$ is model-dependent, whereas $\beta_p$ is not.

---

[32] The Didymos system is not strictly Keplerian owing to the non-spherical shapes of both bodies. However, the expected error in equation (7) resulting from this assumption is only a few percent, which will be a negligible contribution to the total error budget for $\beta_p$ or $\beta_n$.

## 2.2 Beta for a Deflection Mission

A momentum enhancement factor linked to a specific direction will also be useful for an actual asteroid deflection mission, in which there will be a preferred deflection direction $\hat{\boldsymbol{u}}$ that will best move the projected path of the asteroid off the Earth (Conway 2001, Vasile & Colombo 2008). It is most convenient to do the analysis in the b-plane framework. The b-plane is oriented orthogonal to the inbound asymptote of the Earth-bound asteroid orbit, with the geocenter at the origin. The intersection of the inbound asymptote with the b-plane is denoted by $\boldsymbol{b}$, a vector in the b-plane, with $b = |\boldsymbol{b}|$. (See, e.g., Farnocchia et al. (2019) and references therein for a thorough development of b-plane theory.) The objective of our supposed deflection campaign is to shift the orbit of a potential Earth impactor to assure that $b$ falls outside the capture cross-section, i.e., $b > b_\oplus$, where $b_\oplus$ is the distance from the geocenter for a grazing impact. Because of gravitational focusing, $b_\oplus$ is always larger than the radius of the Earth by an amount depending on the Earth encounter velocity of the asteroid. We can model the change in $b$ due to a deflection $\Delta\boldsymbol{v}$ according to

$$\partial b = \boldsymbol{u} \cdot \Delta\boldsymbol{v}, \tag{11}$$

Where $\Delta\boldsymbol{v} = \Delta\boldsymbol{p}/M$ and $M$ is the total mass to be deflected (which includes the possibility of a binary; see Section 2.8). The vector $\boldsymbol{u}$ is the crucial factor for the deflection campaign as it maps the effect of the deflection to the b-plane. We can break $\boldsymbol{u}$ into pieces according to the chain rule as follows:

$$\boldsymbol{u} = \frac{\partial b}{\partial \boldsymbol{v}} = \hat{\boldsymbol{b}} \cdot \frac{\partial \boldsymbol{b}}{\partial \boldsymbol{X}_t} \cdot \frac{\partial \boldsymbol{X}_t}{\partial \boldsymbol{v}}. \tag{12}$$

Here $\boldsymbol{X}_t$ is the heliocentric Cartesian state vector (position and velocity) and $\boldsymbol{v}$ is the asteroid heliocentric velocity at the deflection epoch. The matrix of partial derivatives $\partial \boldsymbol{b}/\partial \boldsymbol{X}_t$ reflects the sensitivity of the mapping from asteroid state to the b-plane. The matrix elements depend only on the circumstances of the Earth encounter, which are fixed and thus cannot be changed as a part of the deflection campaign. The last factor, $\partial \boldsymbol{X}_t/\partial \boldsymbol{v}$, represents the so-called state transition matrix and is determined by the asteroid orbit. It can only be modified through the selection of the deflection epoch. Folding in the formalism of equation (6) we can see how $\beta_u$ plays directly into the effectiveness of the deflection campaign:

$$\partial b = \beta_u \frac{m_s}{M} \boldsymbol{u} \cdot \boldsymbol{v}_s. \tag{13}$$

A full analysis of the inner product $\boldsymbol{u} \cdot \boldsymbol{v}_s$ in equation (13) can be quite complicated, as arbitrary deflections can introduce periodic terms that are especially important for short lead time deflections. However, in the case of long lead time, the preferred direction will nearly coincide with the direction of the asteroid's heliocentric velocity at deflection (Vasile & Colombo 2008). This is because an impulse along the velocity vector maps directly to a change in period, and results in a displacement (between the deflected and undeflected asteroid positions) that grows in time over many orbits. In this approximation $\hat{\boldsymbol{u}} = \hat{\boldsymbol{v}}$, and the effect of the deflection is proportional to the change in mean motion $\Delta n$ and the lead time $T$, so that $\delta b \propto \Delta n T$. Applying the vis-viva integral leads to $\delta b \propto Tv\Delta v$, where $v = |\boldsymbol{v}|$ and $\Delta v = \hat{\boldsymbol{v}} \cdot \Delta\boldsymbol{v}$. Incorporating equation (6) leads to

$$\partial b \propto \beta_u m_s T v v_s \cos\lambda, \tag{14}$$

where $\lambda$ is the angle between $v$ and $\boldsymbol{v}_s$. Each factor in equation (14) provides an avenue for improving the effectiveness of a deflection, and so this formula concisely summarizes how to optimize a deflection attempt. As is well known, the spacecraft mass $m_s$ should be maximized, as should the lead time $T$. The deflection epoch should be close to the asteroid's perihelion to maximize $v$, and the projected spacecraft relative velocity should be optimized by maximizing the velocity of impact and intercepting at a point where the spacecraft and asteroid orbits are nearly tangent so that $|\cos\lambda| \approx 1$. Note that the sign of $\cos\lambda$ is important. It determines whether the deflection will push the asteroid closer or farther from Earth, and is driven by whether the spacecraft is overtaking the asteroid at deflection (so that $\cos\lambda > 0$) or vice versa. It is also possible, though time consuming, to achieve high impact speeds using a spacecraft on a retrograde heliocentric orbit (McInnes 2004, Petropoulos et al. 2007).

Finally, the deflection effort should be designed to maximize $\beta_u$. Due to the necessities of celestial mechanics and limitations of spaceflight hardware, the maximum achievable value for the product $v_s \cos\lambda$ will typically not happen with $|\cos\lambda| = 1$, meaning that the spacecraft momentum will not be directed along $u$. Even so, as we shall see in Section 2.7.2, $\beta_u$ can be increased by targeting an oblique impact and selecting an aimpoint on the asteroid surface so that the net ejecta momentum aligns the total momentum change with $u$. For a spherical asteroid we can think of this as an off center vs. centered impact location, though the detailed shape model and spin state of the asteroid would drive the precise impact targeting (Feldhacker et al. 2017). In such an application, $\beta_u$ functions less as a description of the interaction with the asteroid surface and more as a figure of merit for mission design and operations.

So $\beta_u$ for an actual deflection mission is likely to look very similar to $\beta_p$ for DART. $\beta_u$ would be the fraction (intended to be >1) of the spacecraft momentum delivered in the desired deflection direction that is imparted to the asteroid in that direction. For DART, $\beta_p$ is the fraction of the spacecraft momentum delivered in the direction that changes the orbit period that is imparted in that direction. The related – and yet different – quantity $\beta_n$ is the fraction of spacecraft momentum delivered normal to the local surface at the impact point that is imparted in *that* direction, which may or may not be relevant to mission objectives.

## 2.3 Momentum Transfer and Ejecta Response

All of the momentum transfer efficiencies discussed above are determined by the ejecta response of the asteroid surface to the kinetic impact; and one objective of DART is to refine the physical understanding of that response and the ability to model it accurately. In this section we develop a parametrization of the ejecta response and show how a direct determination of $\beta_p$ from DART measurements may be used to constrain the key parameters.

The momentum imparted to the asteroid $\mathbf{\Delta p}$ is the sum of the momentum delivered by the spacecraft $\boldsymbol{p}_s$ and the negative of (*i.e.,* recoil from) the total asymptotic momentum of the unbound ejecta $\boldsymbol{p}_e$. (Only the momentum carried permanently away from the system contributes to $\boldsymbol{p}_e$ in this accounting.) By equation (3),

$$\boldsymbol{p}_e = (\mathbf{1} - \mathbf{B}) \cdot \boldsymbol{p}_s, \tag{15}$$

where $\mathbf{1}$ is the unit tensor. Just as $\mathbf{B}$ maps the spacecraft momentum into the asteroid's momentum change, $(\mathbf{1} - \mathbf{B})$ maps it into the ejecta momentum. This means that all of the essential aspects of the impact physics and surface properties are contained in $(\mathbf{1} - \mathbf{B})$, as is the predictive power to extend this knowledge to new situations. It is conceptually convenient to split $(\mathbf{1} - \mathbf{B})$ into two factors, according to

$$(\mathbf{1} - \mathbf{B}) = \eta \mathbf{H}, \tag{16}$$

where $\eta$ is a scale factor equal to the ratio of magnitudes, $p_e/p_s$, and $\mathbf{H}$ (capital eta) is a pure rotation matrix that rotates the spacecraft direction $\hat{\boldsymbol{p}}_s$ to the ejecta direction $\hat{\boldsymbol{p}}_e$.

To make the connection with impact experiments and simulations, we imagine an impact at an arbitrary location on the asteroid's surface, and set a local Cartesian coordinate system $(x', y', z')$ with the origin at the point of impact and the positive $z'$ axis pointing back along the spacecraft's incoming path (Figure 1a). The $y'$ axis is coplanar with the surface normal $\hat{\boldsymbol{n}}$ and $z'$, so that $\hat{\boldsymbol{n}}$ lies in the first quadrant of the $(y', z')$ plane, and $x'$ points along the surface, to the right as seen by the spacecraft. The spacecraft impacts the surface at angle of incidence (measured from the normal $i = \cos^{-1}(\hat{\boldsymbol{n}} \cdot \hat{\boldsymbol{p}}_s)$), carrying momentum $p_s$ and kinetic energy $E_s$ in the frame of the asteroid.

The ejecta response consists of a deterministic "homogeneous" part and an unpredictable "random" part. The former, resulting from the average properties of the local surface, assumed to be homogeneous over a range of scales encompassing the impact and crater formation process, consists of a scale factor $\eta(i, p_s, E_s)$ and a counterclockwise rotation around the $x'$ axis by an angle $\pi - \gamma(i, p_s, E_s)$. The more intuitive ejection angle, measured from the normal, is $\alpha \equiv \gamma - i$ (Figure 1a). Ideally, with sufficient understanding of asteroidal materials, we would be able to predict the functions $\eta(i, p_s, E_s)$ and $\alpha(i, p_s, E_s)$[33] for a given surface using hydrodynamic and particle simulations. We do not yet have that level of knowledge, but tests like DART are intended to help develop and validate it.

---

[33] The (presumably weak) functional dependence on impactor momentum and energy, as we have written it here, could equally well be cast as a dependence on impactor mass and velocity. In what follows, we omit the explicit dependence for brevity.

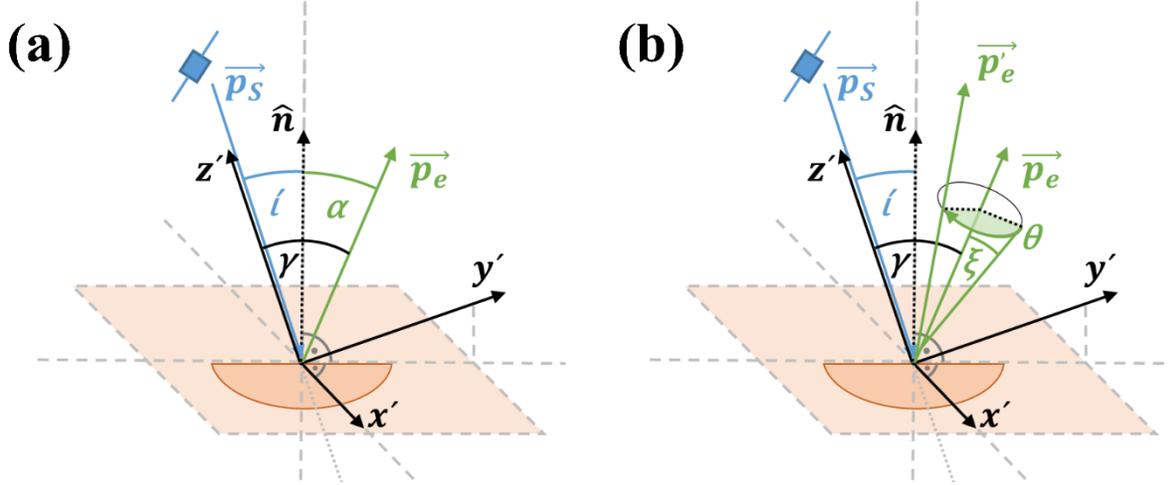

Figure 1: Coordinate system and angles. (a), left: The tan square in perspective indicates the asteroid surface around the point of impact, with the darker semicircle suggesting subsurface material that will be excavated and ejected. The spacecraft arrives traveling in the $-z'$ direction, at an impact angle $i$ from the surface normal $\hat{\boldsymbol{n}}$. The net momentum of the outgoing unbound ejecta, in the homogeneous case, is coplanar with $z'$ and $\hat{\boldsymbol{n}}$, and directed at an ejection angle $\alpha$ from the normal. (b), right: As in (a), showing the parametrization of the random part of the response. The ejecta momentum is increased in magnitude by a factor $(1 + \zeta)$ and altered in direction by an angle $\xi$ at azimuth $\theta$.

The random part of the response results from surface asymmetries or irregularities that are unresolved in any obtainable imaging or hidden beneath the surface. They are intrinsically unknowable, and cause a dispersion in the actual response around the homogeneous prediction. We represent this (Figure 1b) by an angular deviation about the homogeneous direction of magnitude $\xi$ at an orientation $\theta$, where $\xi$ is a random angle with zero mean, and $\theta$ is a second random angle uniformly distributed between $0$ and $\pi$; and by a multiplicative correction $(1 + \zeta)$ to $\eta$, where $\zeta$ is a random dimensionless quantity with zero mean. The distributions of $\xi$ and $\zeta$ may eventually be estimable from simulations but can, for the time being, reasonably be taken to be Gaussians, as their purpose is to estimate uncertainty.

Finally, we establish a coordinate system $(x, y, z)$ for the deflection test or attempt, with the $z$ axis coinciding with $z'$ and the desired deflection direction $\hat{\boldsymbol{u}}$ in the $(y, z)$ plane, and rotate the local picture around the $z$ axis by $\varphi$ so as to put the surface at the impact point in the right orientation in space. The angle $\varphi$ is determined by the asteroid shape model. Putting together the whole sequence of rotations, we have:

$$\mathbf{H} = \begin{bmatrix} \cos\varphi & -\sin\varphi & 0 \\ \sin\varphi & \cos\varphi & 0 \\ 0 & 0 & 1 \end{bmatrix} \begin{bmatrix} 1 & 0 & 0 \\ 0 & -\cos\gamma & -\sin\gamma \\ 0 & \sin\gamma & -\cos\gamma \end{bmatrix} \begin{bmatrix} \cos\theta & -\sin\theta & 0 \\ \sin\theta & \cos\theta & 0 \\ 0 & 0 & 1 \end{bmatrix} \begin{bmatrix} 1 & 0 & 0 \\ 0 & \cos\xi & -\sin\xi \\ 0 & \sin\xi & \cos\xi \end{bmatrix}; \quad (17)$$

and, using equations (6) and (16),

$$\beta_u = 1 + \eta(1 + \zeta)\{\sin\gamma\cos\theta\sin\xi + \cos\gamma\cos\xi \\ + \tan\lambda\left[\cos\varphi\left(\cos\gamma\cos\theta\sin\xi - \sin\gamma\cos\xi\right) + \sin\varphi\sin\theta\sin\xi\right]\}, \quad (18)$$

where $\lambda = -\tan^{-1}(u_y/u_z)$ is the angle between the spacecraft direction $-\hat{\boldsymbol{z}}$ and the desired deflection direction $\hat{\boldsymbol{u}}$. Remember that the homogenous physics of the average surface is contained in $\eta$ and $\gamma$, and the random influence of surface irregularities is contained in $\zeta$, $\xi$, and $\theta$. For the purely homogenous case, $\zeta = \xi = 0$, and equation (18) simplifies to

$$\beta_u = 1 + \eta(\cos\gamma - \tan\lambda\cos\varphi\sin\gamma), \quad (19)$$

which demonstrates straightforwardly that, given the circumstances of the deflection ($\lambda$) and knowledge of the orientation of the impact site ($\varphi, i$), a measurement of $\beta_u$ defines a joint constraint on the parameters ($\eta, \gamma$) describing the ejecta response at that incident angle $i$.

To make the connection with Rivkin et al. (2021) and the vector $\boldsymbol{\epsilon}$, recall that $\boldsymbol{\epsilon}$ is perpendicular to the surface normal $\hat{\boldsymbol{n}}$ and has magnitude $\tan \alpha'$, where $\alpha'$ is the angle between $\hat{\boldsymbol{n}}$ and $\hat{\boldsymbol{p}}_e$. With that definition, the ejecta momentum vector $\boldsymbol{p}_e$ and its normal component $p_{en}$ are related by $\boldsymbol{p}_e = (\hat{\boldsymbol{n}} + \boldsymbol{\epsilon})p_{en}$. In terms of the parametrization of this paper, in the homogenous case, $\alpha' = \alpha \equiv \gamma - i$, and $\boldsymbol{\epsilon}$ points in the downrange direction. Explicitly including the random part of the response yields

$$\alpha' = \cos^{-1}[\cos(\gamma - i) \cos \xi - \sin(\gamma - i) \sin \xi \cos \theta]; \tag{20}$$

however, in this case $\hat{\boldsymbol{p}}_s$, $\hat{\boldsymbol{n}}$, and $\hat{\boldsymbol{p}}_e$ are no longer coplanar. Instead, $\boldsymbol{\epsilon}$ points an azimuthal angle $\delta$ away from the coplanar downrange direction, where

$$\delta = \sin^{-1} \frac{\sin \alpha' \sin \xi}{\sin(\gamma - i)}. \tag{21}$$

## 2.4 Potential Constraints from DART

Now that the DART spacecraft has been launched, the geometry of its kinetic impact on Dimorphos on 2022 September 26 is set, and so it is known that the angle $\lambda$ between the spacecraft direction and the asteroid's orbital direction will be 166°, with a 1-$\sigma$ systematic uncertainty of approximately 3° owing to imprecise knowledge of Dimorphos's orbit pole (Scheirech & Pravec 2022). The pre-impact orbital period is known to extremely high precision from repeated observations of mutual events, and the post-impact period will, for the same reason, become more precise with time. Thus the uncertainty in $\beta_p$ calculated by equation (9) will be dominated by the uncertainties in the total system mass, currently known to approximately 7.5% (Richardson et al. 2022), and in the mass of Dimorphos, which is currently weakly constrained. Following impact, DRACO and LICIACube images will be used to construct a shape model that is expected to constrain the volume of Dimorphos to approximately 30% (Daly et al. 2022); this, with an assumed density (nominally equal to the bulk density of Didymos), will result in a credible mass estimate, albeit one subject to large systematic error. Thus, following DART, we can expect a determination of $\beta_p$ uncertain to < 10% (statistical) plus a factor $\sim 2$ (systematic). This situation will be only temporary, however, since the Hera rendezvous and proximity operations will permit the masses of both Didymos and Dimorphos to be directly measured in situ with much higher accuracy (Michel et al. 2022). Hence, we can reasonably expect that $\beta_p$ will be determined to better than 10%, if not immediately after DART, at least after AIDA.

The Dimorphos shape model (likely to be refined after Hera's arrival) will also be used to determine the orientation of the local surface at the impact point, providing constraints on the angles $i$ and $\varphi$. To illustrate how all of this will constrain the ejecta response, we arbitrarily imagine that it has been found that $i = 30°$, $\varphi = 120°$, and $\beta_p = 2.1 \pm 0.2$. (We assume uncertainties in the angles are small so as not to overcomplicate the illustration.) Figure 2a shows the result of a Monte Carlo simulation in which points have been thrown down randomly, and a priori uniformly, in the ($\eta, \alpha$) plane and tested against the measured $\beta_p$ and uncertainty; the density of points around an arbitrary ($\eta, \alpha$) pair is proportional to the likelihood that the ejecta response described by this ($\eta, \alpha$) reproduces the true value of $\beta_p$. This defines a fuzzy locus in the ($\eta, \alpha$) plane constrained by the experimental results. Figure 2a assumes that the ejecta response is fully homogenous with $\zeta = \xi = 0$; Figure 2b shows a more realistic case including surface irregularities, with values of $\zeta$ and $\xi$ drawn from Gaussian distributions of 1-$\sigma$ width 0.1 and 10°, respectively. Not surprisingly, surface irregularities widen the constraint on the underlying homogeneous values of $\eta$ and $\alpha$.

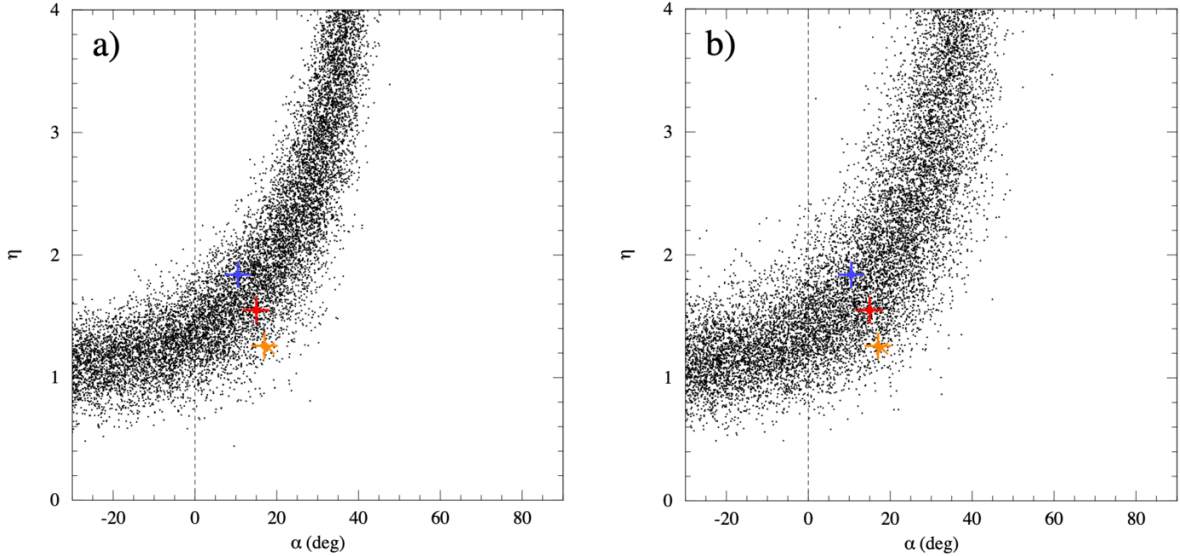

Figure 2: Constraints on the ejecta response $(\eta, \alpha)$ obtained from a hypothetical measurement, $\beta_p = 2.1$, for the DART kinetic impact. Local density of small black points is proportional to the likelihood that the corresponding $(\eta, \alpha)$ pair reproduces the measured $\beta_p$. (a), left: Homogenous case, with no surface irregularities. (b), right: Including the random effect of surface irregularities. Vertical dashed line corresponds to ejecta momentum normal to the surface. Colored points indicate predictions from hydrodynamic simulations, as in Figure 3.

Remember that this is a constraint on the ejecta response at a single incident angle $i$. A set of simulations performed with fixed material properties, and fixed impactor mass and energy, will yield a single pair of $\eta$ and $\alpha$ values at each $i$. With the impactor properties known in advance, as in the case of DART, this allows a determination of whether the assumed material properties are consistent with the experimental result. A grid of oblique impact simulations with varying material parameters would produce a grid of predictions in the $(\eta, \alpha)$ plane, potentially allowing regions of material parameter space to be ruled out. The colored points in Figure 2 provide an example, which we discuss in Section 2.6.2 below.

A single test mission will not be able to constrain the ejecta response at all incidence angles, or for all impact energies or impactor masses; yet, a goal of kinetic impact studies is to develop an accurate physical understanding that spans a range of geometries and surface properties. To achieve that, it will be necessary to rely on physical scaling laws as well as enlarged grids of simulations. Although 3-D hydrodynamic simulations are currently time-consuming and expensive, there is no fundamental obstacle to a comprehensive simulation survey of the relevant multidimensional parameter space; we discuss this further in Section 2.6.4 below. For this the DART test will serve as an initial – and, for the time being, singular – ground-truth anchor.

It should be mentioned that we are neglecting an issue specific to DART. A fraction of the slow ejecta that exceed the escape speed from the surface of Dimorphos, $v_{esc}$, will not exceed the escape speed from the Didymos system (Fahnestock et al. 2022, Rossi et al. 2022). Ejecta exceeding only about $0.75 v_{esc}$ will be launched outside of Dimorphos's Hill sphere. This material may orbit in the system for some time before being gravitationally ejected or impacting either body. It is conceivable that long-lasting fragments could find refuge near Lagrange points or even form a ring, either around the primary object or on circumbinary orbits, which could be detectable by Hera. The effects of late-escaping or re-impacting ejecta on the interpretation of the measured period change are under study by the DART team, and are outside of the scope of this paper.

## 2.5    Additional Constraints from LICIACube and Hera

Further in situ data acquired after the DART impact will serve to refine the understanding of the geophysical and compositional properties of the Didymos system, and will help connect the constraints on the ejecta response obtained from the test to observable geophysical signatures of the underlying material properties.

Images of the ejecta plume from LICIACube will potentially add constraints on the ejecta momentum, especially the ejection angle $\alpha$. Images of the newly-forming crater, if visible through the plume, will provide additional constraints on the near-surface porosity and help to constrain the other ejecta response parameters. The full LICIACube dataset should provide information on surface properties of Dimorphos, such as geological features, color differences and variegation, boulder distribution, and mass movements if present (Pajola et al. 2022, Poggiali et al. 2022), as well as information on the ejecta plume structure and dust dynamics evolution shortly after the DART impact, up to the distance of the LICIACube closest approach (Ivanovski et al. 2022, Rossi et al. 2022).

Hera, beyond the precision measurements of mass, shape, and volume (and consequently bulk density) already discussed, should provide additional information about the surfaces and interiors of both Didymos and Dimorphos (Michel et al. 2022), including:

- The size and morphology of the final DART impact crater. Crater diameter and depth are predicted by hydrodynamic models to correlate with porosity and cohesion.
- The rotation state of Dimorphos, which could be left significantly altered by the impact (Agrusa et al. 2021).
- Possible signs of surface transformations induced by the impact itself, the fallback of debris, or mobilization of regolith (Quillen et al. 2022, Hirabayashi et al. 2022, Agrusa et al. 2022).
- Detection of, or limits on, long-lived orbiting ejecta (Rossi et al. 2022).
- Size-frequency distributions and shapes of natural craters, boulders, and other surface features such as fractures.
- Low-frequency radar measurements (supplemented by gravimetry) of internal geological structures (layers, large voids, sub-aggregates, etc.), and estimation of the average dielectric permittivity (which further constrains composition and porosity).
- Thermal indications of heterogeneities or exogenous materials.

## 2.6    Ejecta Response Models and Material Properties

### 2.6.1    Simple Heuristic Models and Scaling Relations

Point-source scaling relationships (e.g., Housen & Holsapple 2011) can be used to describe the mass-velocity distribution of the ejecta. Cheng et al. (2016) proposed an analytic expression based on the point-source assumptions to estimate the magnitude of the momentum transfer, $\beta_n$, from scaling relationships for vertical impacts into homogeneous targets. The expression was tested by Raducan et al. (2019), who found agreement with numerical simulations within ~10%. The point source scaling approximation predicts a weak dependence on the projectile velocity, $(\beta_n - 1) \propto v_s^{3\mu-1}$, where the dimensionless $\mu$ parameter depends on target properties and lies in the range $1/3 < \mu < 2/3$. In porous materials, typically $\mu = 0.4$, leading to $(\beta_n - 1) \propto v_s^{0.2}$. The point source scaling also predicts that $\beta_n$ is independent of projectile mass, so that impacts of different mass projectiles at the same density, at the same speed, into the same target material, yield the same $\beta_n$ values.

### 2.6.2    Homogeneous Response: Current Knowledge from Simulations

The effects of target mechanical properties on impact processes and momentum enhancement have been investigated in a number of laboratory and numerical studies (e.g., Cintala et al. 2010, Bruck Syal et al. 2016, Chourey et al. 2020, Holsapple & Housen 2012, Jutzi & Michel 2014, Luther et al. 2018, Raducan et al. 2019, Raducan et al. 2020, Stickle et al. 2015, Stickle et al. 2017, Raducan & Jutzi 2022, Luther et al. 2022), a comprehensive summary of which, specifically in the context of DART, is given in Stickle et al. (2022). These studies found that, for vertical impacts, the main determinants of momentum enhancement magnitude are the target material's strength and porosity. An increase in strength, i.e., cohesion and/or internal friction, reduces the resulting crater size and the amount of ejected material. Cohesion has little effect on the ejection speed or angle of high-speed ejecta, whereas increasing the coefficient of internal friction leads to shallower ejection angles (Luther et al. 2018) and less momentum carried in the normal direction to the surface (Raducan et al. 2022). An increase in porosity generally causes a reduction in the crater diameter and the amount of ejected mass, owing in part to the lower density for the same material. But the main effect is the reduction of ejection speeds due to energy dissipation in pore crushing, which leads to lower momentum carried in the ejecta and also shifts the bulk of the momentum contribution toward ejecta of lower speed.

The dependence of the ejecta response on impact angle is an active field of study, both experimentally and numerically (e.g., Pierazzo & Melosh 2000 and references therein, Yanagisawa & Hasegawa 2000, Anderson et al. 2003, Ebelshausen et al. 2009, Shuvalov 2011, Stickle et al. 2015, Raducan et al. 2022). It is generally expected that oblique impacts will reduce the momentum enhancement in the normal direction (e.g., Stickle et al. 2015, Feldhacker et al. 2017, Raducan et al. 2022), although, as we have shown above, this is not necessarily the most relevant direction in a real deflection attempt.

Figure 3 shows results from numerical simulations of DART-like impacts carried out with Bern's grid-free smoothed-particle hydrodynamics (SPH) shock-physics code (Jutzi, Benz, & Michel 2008, Jutzi 2019) to study more deeply the effects of oblique impacts on momentum enhancement and ejecta properties for low-strength materials. We simulated 620 kg spherical projectiles impacting porous basaltic regolith targets at 7 km/s, at various impact angles. The target porosity was kept constant at 20% and was modeled using the P-$\alpha$ model (Jutzi, Benz, & Michel 2008) with a simple quadratic crush curve with the input parameters defined in Luther et al. (2022). The target was modeled as a half-space with a cohesive strength between $Y_0 = 0.1$ and 10 kPa. The material response to shear deformation was described by a simple pressure-dependent strength model (Collins et al. 2004), which asymptotes to a certain strength at high pressures. The resolution of the simulations was about 5 million SPH particles in the target. Also shown for comparison are results using the iSALE-3D code for the case of $Y_0 = 0.1$ kPa (Raducan et al. 2022), which are similar to the corresponding SPH results and illustrate the level of agreement between the two approaches.

Figure 3a shows the ratio of momentum magnitudes $\eta = p_e/p_s$ as a function of impact angle $i$.[34] At a fixed impactor momentum, the magnitude of the net ejecta momentum decreases as the impact becomes more oblique. This is to be expected since oblique impacts result in lower shock pressures (Pierazzo & Melosh 2000), smaller crater volumes (Ebelshausen et al. 2009) and less ejected material, (e.g., Raducan et al. 2022). Figure 3b shows the ejection angle $\alpha$ as a function of $i$. In these simulations, $\alpha$ ranges from 0° for vertical impacts to close to 40° for an oblique impact at $i = 60°$; higher values of $i$ were not run. Finally, Figure 3c plots $\eta$ vs. $\alpha$. As the impact becomes more oblique, the momentum carried by the ejecta in the direction normal to the surface (proportional to $\eta \cos \alpha$) is reduced while the downrange component (proportional to $\eta \sin \alpha$) increases.

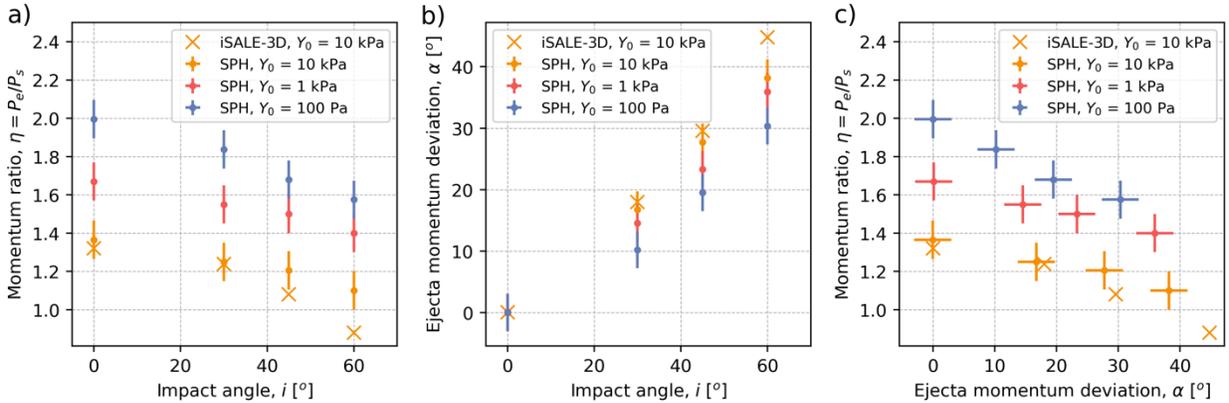

Figure 3: Results from Bern SPH oblique impact simulations (circles with error bars) and from iSALE-3D (Raducan et al. 2022; crosses) on half-space targets. a) Ratio of ejecta momentum to impactor momentum $\eta$ as a function of impact angle $i$, for target cohesions $Y_0 = 0.1$, 1 and 10 kPa. b) Ejection angle $\alpha$ as a function of impact angle $i$. c) Momentum ratio $\eta$ as a function of ejection angle $\alpha$.

While the simulations above demonstrate the angle-dependent ejection response over a relatively narrow range of weak strength values, to date no systematic studies have been carried out over a much wider range of material properties for all impact angles. This is due to the need for computationally expensive three-dimensional hydrodynamic simulations that must be run to relatively long times after the transient crater is formed, when most ejecta have detached from the surface. However, we can compare recent simulation results from impact scenarios with different material strengths, modeled using different numerical approaches, at a fixed impact angle of $i = 45°$ (Figure 4). We use results from Bern SPH simulations reported above and in Raducan & Jutzi (2022), as well as results using iSALE-3D (Wünnemann et al. 2006), reported in Raducan et al. (2022), and CTH (McGlaun et al. 1990, Hertel et al. 1995). Both iSALE-3D and CTH are grid-based codes; in brief, CTH is a 2-step Eulerian finite-difference code with an adaptive mesh refinement capability, while iSALE is an explicit Arbitrary Lagrangian Eulerian (ALE) mesh. Both codes rely on a continuum representation of materials and allow for multiple materials and rheologies with the option to apply strength/damage models. The three different codes were used to simulate ~600 kg projectiles impacting at 6 km/s into homogeneous targets of different geometries and different material mechanical properties.

---

[34] Remember that throughout this paper the impact (or incidence) angle and the ejection angle are both measured down from the normal; in some papers the impact angle is measured up from the horizontal.

The Bern SPH simulations considered: (1) spherical (75 m) targets with cohesions between 0 and 50 Pa and 40% porosity; (2) oblate ($85 \times 85 \times 56$ m) and prolate ($100 \times 65 \times 65$ m) ellipsoidal targets, cohesionless (0 Pa) and 40% porosity (Raducan & Jutzi 2022); and (3) half-space targets with a cohesion between $Y_0 = 100$ and 10 kPa and 20% porosity (reported above). The iSALE-3D simulation used a half-space target, with a cohesion of 10 kPa and 20% porosity (Raducan et al. 2022).

The three shock-physics codes use not only different discretization methods, but also different equations to describe the behavior of materials. For example, the targets in both the Bern SPH and the iSALE-3D simulations used the Tillotson equation of state (EoS) and a pressure-dependent strength and damage model as described in Collins et al. (2004). In CTH, the target material was modeled using the Sesame equation of state (Johnson 1994) and the "elastic-perfectly plastic with Von Mises" yield surface (EPPVM) model, with no damage model. The target cohesion was defined within the EPPVM constitutive model that is native to CTH. The porosity-compaction is described by the $P - \alpha$ model in Bern SPH and in CTH, and by the $\epsilon - \alpha$ model in iSALE. Despite the different constitutive models used by the different codes, benchmarking studies (e.g., Stickle et al. 2020, Luther et al. 2021) show that for fixed impact scenarios, the impact outcome varies by only 10-20%. Here we use the results from different codes to show that the trends seen in our results are consistent, regardless of the different numerical approaches. The targets in both the Bern SPH and the iSALE-3D simulations used the Tillotson equation of state (EoS) and a pressure-dependent strength and damage model as described in Collins et al. (2004).

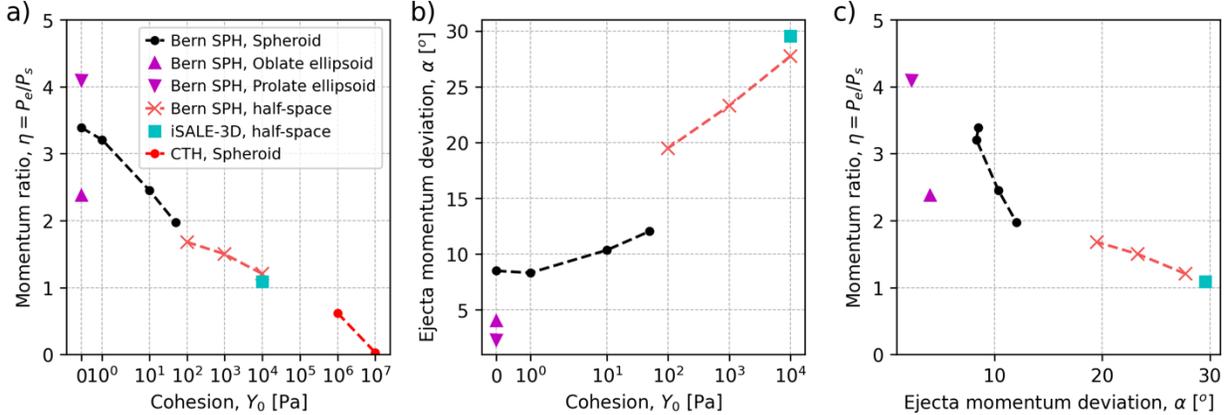

Figure 4: Results from numerical simulations of impacts, at $i = 45°$, from Bern SPH, iSALE-3D and CTH. a) Ratio of the ejecta momentum to the impactor momentum ($\eta$) as a function of target cohesive strength. b) Ejecta momentum direction from the normal, $\alpha$, as a function of cohesion. c) Momentum ratio $\eta$ as a function of ejection angle $\alpha$.

Figure 4a and 4b show, respectively, the ratio of magnitudes $\eta$ and the ejection angle $\alpha$ as functions of cohesive strength, $Y_0$. The same overall trends appear for all simulations. Generally, $\eta$ increases with decreasing cohesive strength, as the crater volume and the mass of ejecta become larger. Also, $\alpha$ decreases with decreasing cohesion, as crater growth takes longer and more material is ejected vertically from the growing crater. There is a temporal aspect to $\alpha$ that is a function of the numerical scheme. This angle should be measured at relatively long time after the transient crater is formed, when most ejecta has detached from the surface (Raducan et al. 2019, Raducan & Jutzi 2022). CTH results are shown only in panel (a) because only a small amount of ejecta was created from the impact into a strong target and $\alpha$ could not be determined to adequate precision. The simulations show a significant difference in $\eta$ between impacts into cohesionless oblate spheroids compared to prolate spheroids with the same mechanical properties. This difference likely arises because of the small size of the targets and the difference in the curvature of the surface at the point of impact. Figure 4c shows $\eta$ vs. $\alpha$, as in Figure 3c. Multiple effects influence $\eta$ at the tens-of-percent level (cohesive strength, tensile strength, internal friction, porosity, etc.), while $\alpha$ appears primarily determined by cohesive strength. Weaker surfaces may therefore be expected to generate less downrange and more vertical ejecta, with lower resulting uncertainty in the predicted deflection, although a lengthy period of large crater formation and ejecta launching could introduce further complications.

As an example of how material parameters may be constrained by the DART test and the determination of $\beta_p$, we replicate in Figure 2 the iSALE and Bern SPH results shown in Figure 3c for the imagined impact angle $i = 30°$. In this fictional but realistic scenario, the measured $\beta_p$ would favor lower values for Dimorphos's near-surface cohesive strength (at the density and porosity assumed in the simulations), demonstrating that meaningful constraints on material physical properties will be obtainable.

### 2.6.3 Effects of Target Inhomogeneities

Crater formation on a heterogeneous, rubble-pile-like target is a complex process, which involves fragmentation, ejection and displacement of boulders. Simulating impacts into heterogeneous targets in the low-strength regime is very challenging (both in laboratory experiments and in numerical simulations). For this reason, there are only a limited number of studies that explore the cratering mechanism in rubble-pile targets and we do not yet know the full range of impact outcomes. Some recent results (Raducan et al. 2022) indicate that there is not very much difference in ejecta momentum magnitude and direction between homogeneous objects and loose-packed rubble piles if there is a distribution of boulder sizes.

Recent laboratory experiments conducted at the Experimental Projectile Impact Chamber (EPIC) at Centro de Astrobiología CSIC-INTA, Spain (Ormö et al. 2022), at low speeds ($\approx 400$ m/s) into targets specifically designed to mimic rubble-pile asteroid surfaces shed some light into possible impact outcomes. It is found that for this target set-up, the cratering event displaces and ejects boulders, rather than fragmenting them, unless the boulders are directly hit. This is consistent with the behavior of the Iijima boulder on Ryugu, which remained intact but was moved by the Hayabusa2 Small Carry-on Impactor (SCI) impact (Arakawa et al. 2020, Honda et al. 2021). The presence of the boulders within the target promotes a steeper ejection curtain angle compared to homogeneous targets. In the EPIC experiments, the ejected boulders segregate from the fine ejecta, resulting in boulders landing at larger distances than the surrounding granular media. At a real kinetic impactor scale and within a low-gravity, vacuum environment, only the very slowest ejecta, gravitationally bound to the target, would be expected to land back on the target, regardless of particle size. However, these ejected boulders can affect the magnitude of the ejecta momentum and its direction.

The number and momenta of the boulders ejected from the surface, as well as the magnitude and direction of the ejecta momentum vector, will depend on the initial boulder distribution on the asteroid's surface. However, systematic studies are needed in order to quantitatively determine the sensitivity of the impact outcome to the boulder configuration. Further studies will help improve our current understanding on the effects of boulders at the scale of a kinetic impactor.

### 2.6.4 Pathway to Improved Models and Parameter Space Coverage

The preceding discussion should make it evident that current understanding of the ejecta response of asteroid surfaces to impacts is incomplete. In principle, a campaign of numerical simulations (anchored by laboratory experiments) could be conducted to fully explore the relevant parameter spaces and computational approaches. In practice, a comprehensive description of ejecta dynamics requires coupling between distinct phases dominated by different physical processes (e.g., Ferrari et al. 2022). Applying different codes to different regimes of validity, handing off from one code to the next, may be a viable strategy; decisions as to which approach is best suited for each phase may be informed by the physical parameters of the target and the near-field environment (Ivanovski et al. 2022). Nonetheless, simulation from first principles is still both conceptually challenging and prohibitively time consuming.

Two complementary approaches have been designed to solve this problem, both of which aim to generalize the prediction of a collision from a small dataset of sample simulations. The first is to fit a scaling law to the dataset; that is, an analytic relationship between impact properties and the outcome for any collision in a physical regime, assuming invariance with respect to one property, usually the mass of the target (e.g., Benz & Asphaugh 1999,

Housen & Holsapple 2011, Cheng et al. 2016). The second is to use the dataset of simulations to train a machine-learning function to predict the outcome of any collision within a known level of accuracy with respect to the "parent" 3-D hydrocode, but with a much shorter runtime (e.g., Cambioni et al. 2019, Valencia et al. 2019, Emsenhuber et al. 2020, Timpe et al. 2020). The scaling-law approach provides insights into collision physics; on the other hand, the machine-learning approach is fully data-driven and provides the parametric function (e.g., neural network) that least overfits the data. Physics inference can then be performed a posteriori by sampling the parameter space at a very high resolution and looking for trends in the outcome (Cambioni et al. 2019).

## 2.7   Informing a Future Planetary Defense Mission

We now fast-forward to the hypothetical future situation in which a potential Earth-impacting object has been discovered and one or more deflection missions must be planned. We assume that the discovery has been made several years before the potential catastrophe, and that the size of the object is such that a kinetic impact is the preferred approach. How would the information gained from DART and Hera inform mission design?

### 2.7.1   Target Characterization

Once a credible non-zero probability of Earth impact were determined, focused characterization efforts would commence, in parallel with astrometric observations to refine the orbit and the impact hazard prediction. Obtaining constraints on mass, size, and shape would be paramount. Information on geological properties would also be important, both for constraining the bulk density (and hence mass), as well as for constraining material properties and the ejecta response. Insights into the porosity and strength of surface regolith and boulders and the interior porosity and coherence would be especially valuable.

We consider three scenarios for the type of characterization that could be achievable: (i) Earth-based observations only; (ii) Earth-based observations plus a rapid-response flyby; and (iii) Earth-based observations plus a Hera-like rendezvous mission. Which scenarios are feasible would depend on the time until impact, since the time needed to execute increases from scenario (i) to (iii).

**(i) Earth-based observations only:** Optical photometric observations would constrain the spin state and shape of the asteroid, as well as provide albedo estimates, which would constrain the size. Observations in the thermal IR from NEO Surveyor could yield a mean diameter accurate to 10% or better. High-accuracy diameters, as well as shape constraints, can be obtained from stellar occultation observations (e.g., Buie et al. 2015), which also provide precision astrometry (Ferreira et al. 2022) and serve to refine the orbit. If the object were to have an Earth close approach nearer than $d_{lim}$, where

$$d_{lim} \approx 0.06 \left(\frac{D}{100 \text{ m}}\right)^{1/2} \left(\frac{\hat{\sigma}}{0.1}\right)^{1/4} \text{AU},
\qquad (22)$$

with $D$ being the object's diameter and $\hat{\sigma}$ its radar albedo, then a ground-based radar system having the capabilities of the legacy Arecibo Telescope could be used to constrain size, shape, spin, surface properties, and to discover or confirm natural satellites (e.g., Benner et al. 2015, Brozovic et al. 2017). For the Goldstone Solar System Radar (DSS-14), currently the world's most powerful planetary radar, the limiting distance would be approximately a factor of 2 smaller. Also, equation (22) assumes a rotation period of 2.1 hr; faster rotation would further decrease the limiting distance. Radar-derived volumes can be accurate to within a few percent (Nolan et al. 2013, Barnouin et al. 2019, Daly et al. 2020) if viewing geometries are optimal. Observation of a satellite could enable determination of the mass of the system and its mean density. Without a satellite, if the semimajor axis drift caused by the Yarkovsky effect is measurable, mass estimates within a few percent could be possible (Chesley et al. 2014, Scheeres et al. 2019). These data would also collectively constrain density and bulk porosity. Near-Earth asteroids (NEAs) approaching close to Earth may also be detectable, or even resolvable, by the Atacama Large Millimeter/Submillimeter Array (ALMA; Lovell 2008). ALMA observations of thermal emission potentially allows measurements of composition, grain size, and porosity complementary to those obtained at other wavelengths (de Kleer et al. 2021, Cambioni et al 2022).

The 0.45 – 2.4 μm reflectance spectrum defines an asteroid's spectral type (e.g., Tholen 1989, Tholen & Barucci 1989, DeMeo & Carry 2013) Although spectra primarily constrain composition, and composition by itself does not reveal physical morphology, spectral classification, tied to past data from spacecraft (Fujiwara et al. 2006, Huang et al. 2013, Lauretta et al. 2019, Veverka et al. 2000, Watanabe et al. 2019), can provide at least some insight into geological properties. For example, S-type asteroids are expected to possess a stony-rich evolved composition that is siliceous in nature and includes olivines, pyroxenes and Fe-Ni metal (Bus & Binzel 2002a, 2002b). This type of asteroid would be expected to possess coarse-grained (mm to cm sized) regolith and possibly smooth terrains as observed on Itokawa and Eros (e.g., Barnouin-Jha et al. 2008, Miyamoto et al. 2007, Veverka et al. 2001, Yano et al. 2006). B- and C-type asteroids, if top-shaped like Bennu or Ryugu, should possess an intimately mixed surface of both coarse regolith and boulders (DellaGiustina et al. 2019, Lauretta et al. 2019, Sugita et al. 2019). Although bulk porosities of S-type asteroids, inferred from comparison of their bulk densities with those of meteorite analogues, vary from 25% for Eros (Wilkison et al. 2002) to 45% for Itokawa (Abe et al. 2006), individual surface rocks on Itokawa are not particularly porous (Tsuchiyama et al. 2011). By contrast, boulders on the B-type Bennu and C-type Ryugu have micro-porosity of approximately 40% to 50% (Scheeres et al. 2015, Rozitis et al. 2020, Yada et al. 2022, Pilorget et al. 2022), and may be fairly weak (Ballouz et al. 2020, Rozitis et al. 2020), with tensile strength about a factor of 10 smaller than most terrestrial rocks. The cohesive strength of the surface regolith of B- and C-types is inferred to be well below 100 Pa (Arakawa et al.,2020), probably closer to 0.6 Pa (Barnouin et al. 2022a). B- and C-type asteroids also show evidence for internal stiffness (Barnouin et al. 2019, Daly et al. 2020, Hirabayashi et al. 2020, Watanabe et al. 2019), and may have large cores with macro-porosities under 25% (Yada et al. 2022. Their bulk porosities are likely large (~50%), as evinced by asteroid Mathilde (Yeomans et al. 1997).

Information on other spectral types is limited. Rosetta obtained low-resolution images of the E-type asteroid Šteins (Besse et al. 2012, Jorda et al. 2012). Likely composed of Fe-O free enstatite, with minor albite plagioclase (Keil 2010), this asteroid appears to be well cratered, with smooth areas. Other asteroids visited by spacecraft such as Lutetia, Vesta and Ceres are significantly larger and are not well suited for inferring the properties of small NEAs.

**(ii) Earth-based observations plus rapid-response flyby:** Earth-based observations may be insufficient to provide even the essential physical information. If a timely measurement of Yarkovsky drift were impossible, if the object did not come within range for radar, and/or if a space-based IR capability were not in place, the mass could be uncertain by an order of magnitude or more. Moreover, because many NEA types have not been visited by spacecraft, inferring surface properties from telescopic data for many newly discovered objects would be fraught with uncertainty (Daly et al. 2022).

A rapid-response flyby could aid in partially filling knowledge gaps. Direct gravity measurements using the Doppler shift of the received radio signal may be achievable for larger objects. The proposed "optical gravity" technique (Bull et al. 2021), in which the main spacecraft would optically track two or more deployed CubeSats that pass the asteroid on different trajectories, also shows promise. A gravity measurement combined with accurate navigation knowledge would then yield an accurate mass estimate. Imaging during the flyby would also allow refinements to the shape model and potentially reduce volume errors to <30% (Daly et al. 2022) and possibly as low as 10% (Sierks et al. 2011), which, with the mass, would constrain bulk density and porosity. High spatial resolution (< m/pixel) surface observations may aid in constraining regolith strength (Barnouin et al. 2022a, Jawin et al. 2020, Perry et al. 2022, Arakawa et al. 2020) and boulder strength (e.g., Ballouz et al. 2020, Rozitis et al. 2020), as well as their distributions across the surface, although a direct interaction is the only sure way to measure the response of the surface, e.g., (Lauretta et al. 2022, Walsh et al. 2022). The presence of very large boulders that could not be the product of an observed surface crater, e.g. (Abe et al. 2006) or color observations that show evidence for heterogeneous properties (DellaGiustina et al. 2020, Tatsumi et al. 2020) could aid in establishing whether or not the target is a rubble pile. Itokawa, for example, was found by Hayabusa to show a considerable amount of heterogeneous regolith, distributed non-uniformly, and exhibiting different mechanical properties resulting from the effects of shock and microfracturing (Fujiwara et al. 2006, Tanbakouei et al. 2019). Some insights on interior strength may be achieved with asteroid shape assessments (Barnouin, et al., 2019; Daly, et al., 2020; Hirabayashi, et al., 2020), lineament evaluations (Besse et al. 2014, Buczkowski et al. 2008) and, if present, evidence for other surface process such as mass movements (Hirabayashi et al. 2020, Walsh et al. 2008, Scheeres 2015, Sánchez & Scheeres 2018, Zhang et al. 2022. However, the short duration of the flyby and the resulting limitation on surface coverage could hinder some of these efforts.

**(iii) Earth-based observations plus rendezvous mission:** A rendezvous mission with radio-science and imaging capabilities would essentially guarantee accurate determinations of mass, volume, and shape, and along with them, bulk density and porosity. In addition, complete mapping of surface features (Pajola et al. 2022) and a precise determination of the rotation state would likely be achievable. By tying evidence for surface mass movements to surface slopes, geotechnical stability analyses could constrain the surface cohesion (Barnouin et al. 2022a, Barnouin et al. 2022b) along with analyses of observed ejecta (Perry et al. 2022). Studies of lineaments might provide a sense of interior coherence and strength (Marchi et al. 2015) and references therein. Boulder and crater size-frequency distributions may also provide clues to the surface porosity and strength of the surface.

Additional information would depend on details of the science payload. Imaging spectroscopy in the thermal IR would constrain thermal inertia, which can be used as an indicator of regolith grain size distributions and boulder porosities. Seismic and radar data of the near-surface and interior would be highly valuable. Low-frequency radar, as well as detailed gravity measurements, would allow the deep interior to be probed, revealing the aggregate structure and the organization and sizes of the constitutive materials. Higher frequencies could image the first tens of meters of the regolith down to decimetric scales providing context for remote sensing or spacecraft interaction with the body (Herique et al. 2018). Seismometers placed on the surface could measure the ground deformation due to seismic waves excited either by natural or artificial sources, potentially revealing sub-surface layering and heterogeneities (Murdoch et al. 2017). Enhanced reconnaissance efforts, where direct surface interactions are undertaken using impact (Arakawa et al. 2020), explosives, thrusters (Lauretta et al. 2022), or low-speed penetrometers (Sunday et al. 2022) could give complementary estimates of surface strength, density, and porosity.

### 2.7.2 Deflection Effectiveness

We assume that sufficient time has elapsed between DART and our hypothetical scenario that simulation and laboratory experiment capabilities have advanced, and that by this point we have improved ability to predict the ejecta response over a range of kinetic impactor and asteroid surface properties, using the constraints derived from DART. We expect, once the potential for Earth impact is recognized, that a variety of mission trajectories to intercept the object for flyby, rendezvous, or kinetic impact will be calculated using standard techniques. Each kinetic impactor trajectory will predict the relative velocity vector at impact as well as the optimum deflection direction for mitigating the danger to Earth. Starting from the premise that this work has been completed, and that the angle $\lambda$ between the spacecraft direction and the optimal deflection direction is known for a given trajectory, we consider what might be predicted for the deflection attempt.

For illustration, we make use of two analytic ejecta response models that roughly bracket the results shown in Section 2.6.2, in order to show how differences in the ejecta response might affect mission requirements. Model 1 uses the ansatz adopted by Feldhacker et al. (2017), that the ejecta momentum is always normal to the surface and $\beta_n$ is independent of $i$. For ease of comparison with Model 2 (see immediately below), we set $\beta_n = 2.32$ to match the normal-incidence case of the iSALE-3D simulations for $Y_0 = 10$ kPa reported by Raducan et al. (2022) (orange crosses in Figure 3):

$$\eta(i) = 1.32 \cos i, \qquad \gamma(i) = i. \tag{23}$$

Model 2 follows from the same iSALE-3D simulations, which show that for increasingly oblique impacts the ejecta momentum has a significant and increasing downrange component; we employ polynomial fits to the data in their Table 2, obtaining

$$\eta(i) = 1.32 - 0.0013i - 0.00000165i^3, \qquad \gamma(i) = 1.45i + 0.005i^2, \tag{24}$$

where $i$ and $\gamma$ are in degrees. The tabulated results extend only to $i = 60°$, and the extrapolation to higher angles is highly uncertain. The coefficients in the expression for $\eta(i)$ in equation (24) are chosen so that $\eta(i)$ goes to zero at $i = 90°$, as it does in equation (23). For clarity in the examples below, we are not including the random effects of surface irregularities, although they could be included straightforwardly to gauge their contribution to the overall uncertainty.

In the limit of minimal information, where we knew nothing about the shape of the object, mission planning based on the assumption of a spherical body would be prudent, and this idealized case is illustrative. Figure 5a shows a map of $\beta_u$ over the hemisphere of an idealized spherical asteroid seen by a kinetic impactor on terminal approach. We imagine that the optimal deflection direction $\hat{\mathbf{u}}$ is into the page and 40° to the right ($\lambda = 40°$). The ejection response is described by Model 1, and values of $\beta_u$ are indicated by the color bar to the left. Figure 5b shows the corresponding map for Model 2, on the same color scale. Several points are worth noting:

1. $\beta_u$ is not constant over the impact hemisphere, even if $\beta_n$ is. Given the desire in this example to deflect the asteroid to the right, it is clearly advantageous for the kinetic impact to occur to the left of center, so that ejecta recoil can push in the desired direction.

2. Downrange ejecta matter. Raducan et al. (2022) find that $\beta_n$ is nearly constant in their simulations, which is reproduced by Model 2. If $\beta_n$ were the only relevant quantity, Figure 5a and Figure 5b would be nearly identical, which is clearly not the case.

3. The maximum efficiency $\beta_u$ can be larger than $\beta_n$. This is because all betas are ratios of vector components in a given direction. For normal incidence and normal ejection ($i = \gamma = 0$), $\beta_u = \beta_n = 1 + \eta(0)$; hence in Figure 5, the centers of both projected disks have $\beta_u = \beta_n = 2.32$, and the maximum values to the left are higher (2.52 and 2.65 for Model 1 and 2, respectively).

4. The regions of high efficiency are not particularly large, and downrange ejecta make a significant difference. Blue, violet, and darker colors indicate regions in which $\beta_u < 1$, where the ejecta are working against the kinetic impactor and lessening its effectiveness. These areas occupy approximately 12% of the projected target disk in Figure 5a, and 40% in Figure 5b. For larger angles $\lambda$, Model 2 gives rise to regions of *negative* $\beta_u$, where ejecta actually push the asteroid the wrong way.

5. For spherical objects having the same ejecta response everywhere on the surface, the value of $\beta_u$ averaged over the projected disk can be shown to be independent of direction $\hat{u}$. For Model 1 the disk-averaged $\beta_u$ is identically equal to $1 + \eta(0)/2 = 1.66$; for Model 2 we find it numerically to be 1.33.

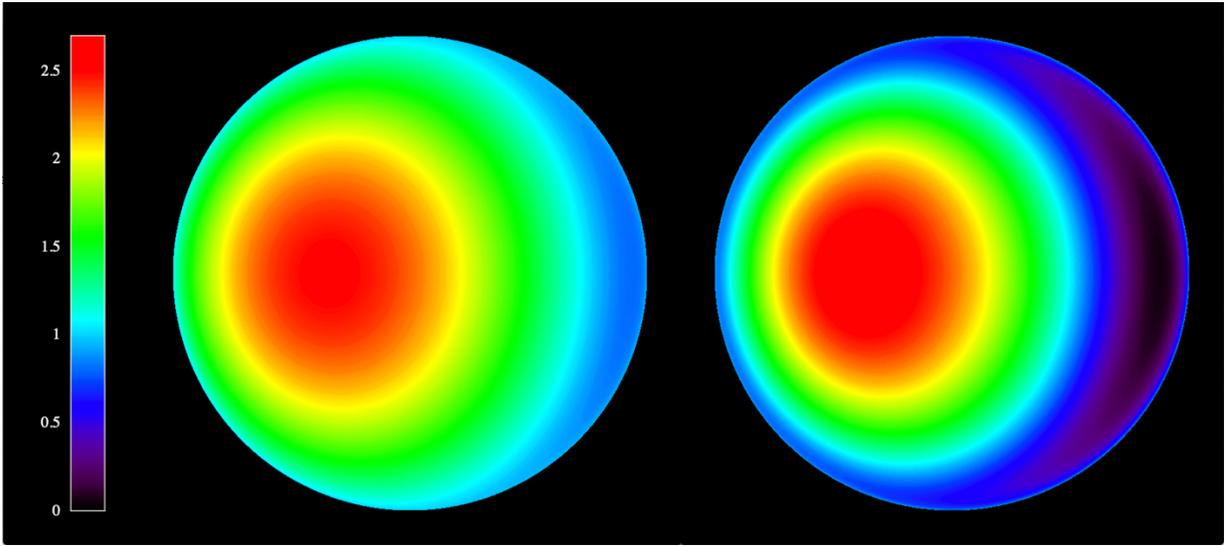

Figure 5: Spacecraft's view of an idealized spherical asteroid showing a map of $\beta_u$ across the surface. Values are indicated by the color scale at far left. The desired deflection direction $\hat{u}$ is into the page and 40° to the right. (a), left: ejecta response Model 1, equation (23); (b), right: ejecta response Model 2, equation (24).

The low values of disk-averaged $\beta_u$ and the potential for sub-unity and negative local values argue for the necessity of precision-targeting an optimal spot on the visible body, so as not to squander the effort. Merely "hitting anywhere" would not be good enough, as the expected result of "anywhere" may be only half as effective as the best achievable, and a "hit" in the wrong spot could even be counterproductive.

In a real planetary defense exigency, it is likely that we will know something about the shape of the object, either from Earth-based observations or from reconnaissance spacecraft. As an illustration of what might be predictable in the limit of *maximal* information, we show in Figure 6 representative maps of $\beta_u$ for an object having the shape of Bennu. We use a shape model (Daly et al. 2020) with a resolution of approximately 6 m, to correspond with a likely size for a kinetic impactor spacecraft. The deflection angle and color scale are the same as in Figure 5, and Figure 6a and Figure 6b show results for ejection response Models 1 and 2, respectively. The Bennu shape model has been put into an arbitrary but not-unreasonable orientation for an imagined deflection near its equatorial plane. The results are reminiscent of the spherical case, but noticeably different in two ways. First, even though the maximum values of $\beta_u$ are the same as for the corresponding spheres, the disk-averaged means are lower, 1.59 and 1.21 for Model 1 and 2, respectively. (Comparable fractions of the disk-projected area – 10% and 44% – have $\beta_u < 1$.) Second, the optimal region for efficient deflection is smaller and not simply connected, making precision navigation in the terminal phase even more essential.

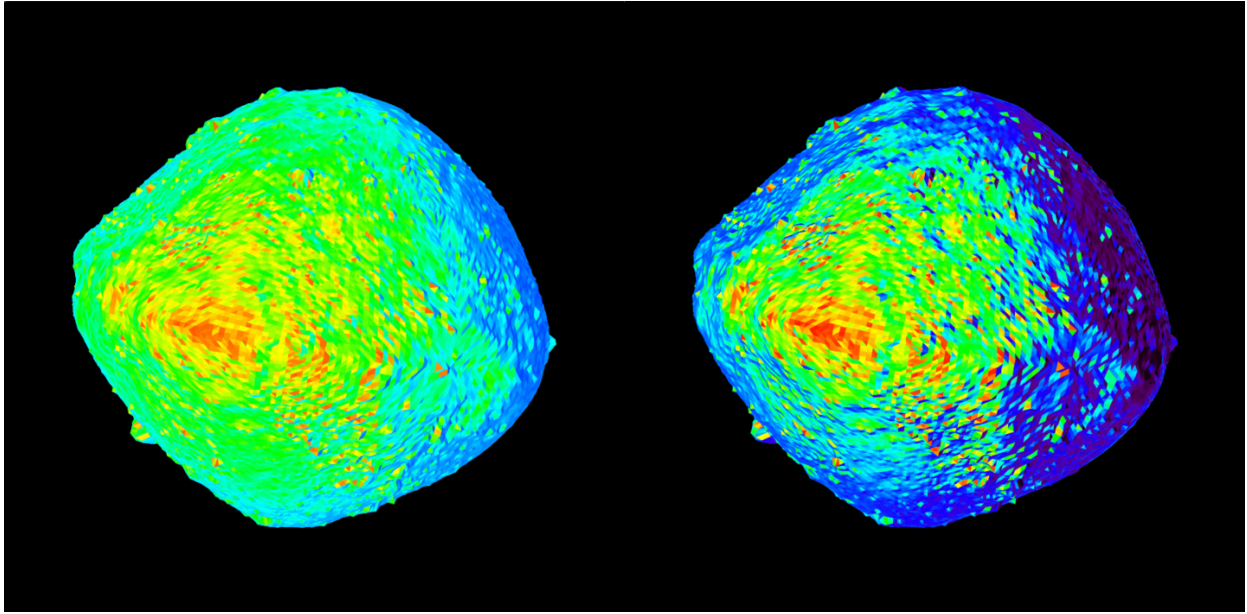

Figure 6: Maps of $\beta_u$, as in Figure 5, for one possible orientation of a Bennu-shaped asteroid. Geometry, ejecta response models, and color scale are the same as in Figure 5.

### 2.7.3    Effects of Rotation

All asteroids rotate, and there are two distinct consequences of rotation for a kinetic impact deflection. The first is the uncertainty in the orientation of the asteroid at impact. In the case of long lead time, a substantial baseline of ground-based light curves could be built up and linked to a shape model further constrained by reconnaissance mission data and/or radar observations, potentially making this uncertainty small or negligible. On the other hand, NEOs with long synodic periods can present infrequent opportunities for light-curve observations, and a deflection may need to occur between these opportunities. In this case, even if the shape were known, the rotation phase at the time of impact might not be, and the best prediction of the deflection efficiency might be a rotationally averaged $\beta_u$ map, resembling the spherical examples in Figure 5.

The second consequence of rotation for kinetic impact is dynamical. We have been assuming that the momentum transfer process takes place over a timescale much shorter than the rotation period of the target. Observationally, the overwhelming majority of asteroids larger than about 200 m in diameter rotate with a period longer than approximately 2 hr (Harris 1996; Pravec et al. 2002), and Dimorphos is assumed to rotate synchronously with its 11.92 hr orbit (Rivkin et al. 2021). If crater excavation and ejecta launching end within a few minutes of impact, an assumption of fixed orientation is reasonably well justified. However, the Hayabusa2 SCI experiment, which induced a relatively small impact event on asteroid Ryugu, showed that excavation may continue for many minutes in the gravity regime (Arakawa et al. 2020). For a larger, sub-catastrophic, impact on a similar surface, ejecta launching may possibly continue for hours (Jutzi 2019, Raducan & Jutzi 2022). Furthermore, smaller asteroids rotate more rapidly; nearly all NEAs smaller than 60 m in diameter have spin periods shorter than 2 hr (Statler et al. 2013, Hatch & Wiegert 2015). Many have periods of only a few minutes or less, putting their surfaces under centrifugal tension, so that crater formation would not be in the gravity regime. (The first such "superfast rotator" discovered, 1998 KY$_{26}$ (Ostro et al. 1999), is a rendezvous target for the Hayabusa2 extended mission (Hirabayashi et al. 2021).) The surface material properties of superfast rotators are not well constrained. A preferred explanation for their resistance to mechanical failure and break-up is simply that they are monolithic. However, internal stresses in such asteroids are not particularly high, only tens of Pa (Hirabayashi et al. 2021), leaving open the possibility that they are rubble piles with low cohesive strength (Holsapple 2007; but cf. Harris 2013). Whether such bodies might be more vulnerable to disruption by kinetic impact is an open question.

Rotation can dynamically affect momentum transfer in multiple ways. First, the angle of impact $i$ is changed by classical aberration (i.e., velocity vector addition: away from the normal on the approaching hemisphere, and toward the normal on the receding hemisphere), although for impact speeds in the range of km/s and rotation speeds in the range of cm/s or m/s, the effect on $i$ as well as on the impact velocity is negligible. The corresponding effect on *ejecta* velocities is not negligible, however, as the low-speed ejecta will generally carry most of the momentum (Holsapple & Housen 2012). More slow ejecta launched from the surface in the direction of rotation will be above the escape speed than ejecta launched in the opposite direction. These effects will change both the magnitude and direction of $\boldsymbol{p}_e$. In addition, rotation reduces the magnitude and changes the direction of local gravity on the pre-impact surface, and Coriolis forces may affect crater excavation mechanics. And finally, for low strength materials and/or fast rotation, the ejecta process may take a sufficiently long time that the body orientation changes significantly while it is happening.

Quantitative analyses of these effects are beyond the scope of this paper, but in the long run may be important to prepare for a real planetary defense scenario. Early characterization of the target asteroid's rotation state – at minimum, its rotation rate – could be influential, even in determining the correct strategy for deflection.

## 2.8    Deflecting a Binary Asteroid

In addition to the Didymos system, approximately 16% of NEAs 200 m in diameter or larger are likely to be binaries (Margot et al. 2002); and so a planetary defense scenario involving a binary Earth impactor requires consideration. Imparting a $\Delta v$ to one body sufficient to deflect the bound system by the necessary amount could be problematic, as unbinding the pair or fragmenting the target body could produce undesired results. If approaching at high relative speed, from a guidance-navigation-control (GNC) performance perspective it may be more practical to target the larger primary; and it may be desirable to maximize the dynamical contribution of ejecta to the deflection. Consideration of these factors may inform the choice of where to apply the deflection $\Delta v$.

To assess some of these issues, we use a set of 24,000 simulated Earth-impacting asteroids described in Chesley et al. (2019), which span the combinations of Earth-impacting asteroid orbit parameters anticipated to be present in the population. We then apply the approximate equations in Hernandez et al. (2014) to calculate the $\Delta v$ needed to deflect each simulated asteroid as a function of the time of deflection prior to Earth encounter, location on the asteroid's orbit at which deflection is applied (e.g., perihelion, aphelion, etc.), asteroid orbital elements, and Earth gravitational capture radius. From this ensemble, we calculate the 95[th] percentile orbit-averaged required $\Delta v$ as a function of the lead-time for deflection, and plot this as the black curve in Figure 7. I.e., for a given lead time, 5% of cases would be expected to require a $\Delta v$ above the curve. Comparable studies using different assumptions have arrived at similar curves (e.g., Figure 2 of Sanchez, Vasile, & Radice 2010).

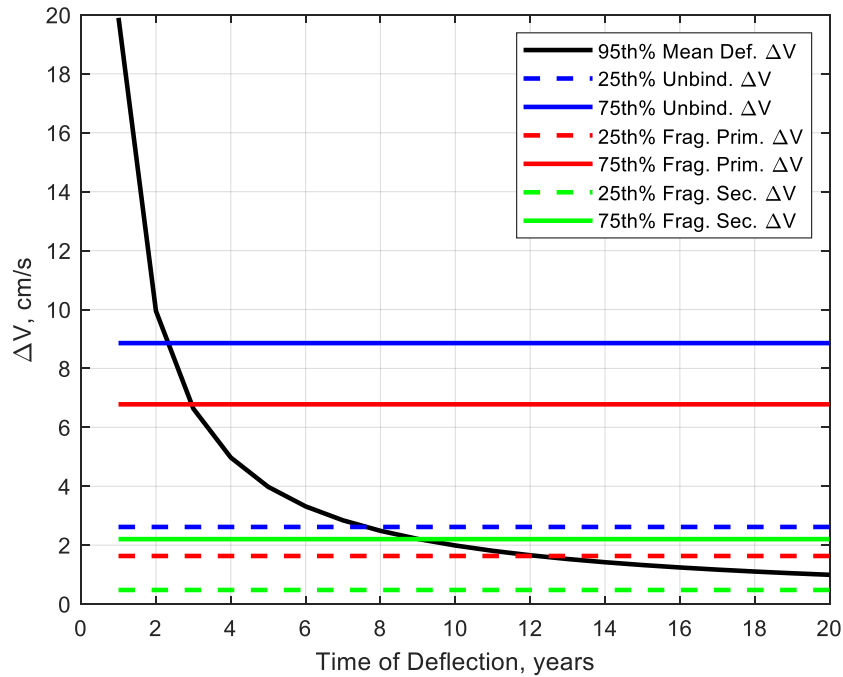

Figure 7: $\Delta v$ required to deflect a large set of simulated Earth-impacting asteroids at the 95[th] percentile level as a function of time of deflection prior to Earth encounter. The horizontal lines show the $\Delta v$, at the 25[th] and 75[th]

percentile levels, that would unbind currently known binary asteroid systems or pose a risk of fragmenting the primary or secondary body in the binary asteroid system.

From the orbital elements and primary and secondary body diameters for the currently known binary NEAs[35], we calculate the minimum $\Delta v$ needed to gravitationally unbind the secondary from the primary in each system (i.e., assuming a "rear-end" kinetic impact that imparts momentum in the direction of the orbital motion, maximizing the orbital energy increase). The 25th and 75th percentiles for this quantity are shown as the blue dashed and solid lines, respectively, in Figure 7. Additionally, the minimum $\Delta v$ that would pose a risk of fragmenting the primary or secondary body is calculated for each pair. Here we assume that a $\Delta v$ equal to 10% of the asteroid's surface escape speed risks fragmentation; this assumption has been used in hypothetical emergency response scenarios[36] but should be examined for accuracy in future work. This is distinct from fragmentation criteria derived from work on strength-dominated materials, given in terms of the impact energy per unit target mass $Q^*$, that indicate threshold values of $Q^*$ in the range of 100 to 1000 J/kg (Sanchez, Vasile, & Radice 2010). Our criterion is intended to apply (more conservatively) to gravity-dominated rubble piles. For Didymos-like target densities and DART-like impactor masses, our criterion corresponds to values of $Q^*$ around the lower end of the cited range. The 25th and 75th percentile values of $\Delta v$ for fragmenting the primary are shown as the dashed and solid red lines, and for fragmenting the secondary as the dashed and solid green lines, respectively, in Figure 7.

A key take-away from Figure 7 is that gravitationally unbinding the pair will typically not be a central concern for a deflection attempt on a binary, because the thresholds for fragmenting the target body are likely to be significantly lower. (And with sufficient knowledge of the system one could choose to target a "head-on" impact as DART does, ensuring that the binary becomes more tightly bound.) For times of deflection less than approximately 8 years, trying to achieve the necessary $\Delta v$ with a single kinetic impactor would be very likely to fragment the secondary if it were the target, and would begin to run a significant risk of fragmenting the primary if it were instead. For times of deflection between 10 and 20 years, the risk of primary fragmentation becomes relatively low while the risk of secondary fragmentation remains moderate even if the deflection 20 is years before Earth encounter.

Minimizing the amount of ejecta that remains bound in the binary system may improve overall deflection performance. If both bodies have similar mechanical properties, ejecta binding to the system is a function of the bodies' locations in the system's gravity well. The escape speed for ejecta from the system can be approximated by summing the effects of gravity from both bodies at the surface of the target (Makadia et al. 2022). A simple analysis shows that the secondary is the better target when the ratio of secondary to primary radii is greater than the corresponding ratio of masses. Only if the secondary has substantially higher density than the primary can the latter become the better choice based on ejecta dynamics alone. Of course, if the secondary is small enough that the risk of unwanted fragmentation is deemed too great, or GNC challenges give rise to a substantial risk of missing, then targeting the primary may be preferred.

## 3    DISCUSSION

Strictly speaking, the pathway from the results of DART to the validation of impact simulation techniques, to constraints on material physical properties, and thence to future planetary defense missions does not have to pass through an interim quantity like a $\beta$. Momentum transfer efficiencies are useful conceptual tools; however, we have argued that the normal-component $\beta_n$ is not by itself a sufficient description of a kinetic impact event, because the ejecta response to the impact is a vector quantity. All components of the ejecta momentum are relevant because the optimal direction in which to deflect a threatening asteroid is set by orbital mechanics, not by the orientation of the surface. At the same time, theoretical understanding of how asteroid surfaces respond to impacts in general should be situationally independent. The normal-component $\beta_n$ and the direction-of-interest $\beta_u$ are distinct, but related and *complementary* quantities. $\beta_n$ answers the question, "What happened – or will happen – at the impact site?" $\beta_u$ answers the question, "How efficiently did we – or can we – deflect the asteroid?"

---

The broad range of possible outcomes implied by the $\beta_u$ maps in Figure 5 and Figure 6 argues in favor of a rendezvous reconnaissance mission that would arrive before any impactors. Even though one might legitimately question why resources and time should be devoted to a rendezvous spacecraft when the same effort could be invested in an impactor that is technically simpler, if there is substantial gain to be had by using the impact ejecta to optimally steer the deflection, or if there is substantial risk of impacting in the wrong place and either being ineffective or making matters worse, than a rendezvous could be well worth the cost. With a reconnaissance spacecraft in place, it could be possible to target regions of the asteroid that may produce strong enhanced momentum, such as broad areal extents of regolith where more material is more likely to readily excavated and bigger thrust generated. (Abe et al. 2006). One could consider precision timing of the impact for a particular rotation phase, or even laying down a target marker to achieve maximum efficiency. Above all, these considerations argue for ample lead time, and the need to find hazardous asteroids "before they find us" (Yeomans 2016). Follow-up assessment observations will be equally critical, both to determine whether the deflection had the desired effect and to monitor the behavior of the asteroid. Exposure of sub-surface materials could lead to a post-deflection increase in surface activity, resulting in non-gravitational perturbations that could influence the subsequent evolution of the orbit.

Recent in situ measurements of rubble-pile asteroid surfaces suggest that models developed for terrestrial applications may not be directly scalable to the asteroid problem (Arakawa et al. 2020, Ballouz et al. 2020). This is due to the radically different environment produced by vacuum and low-gravity conditions (Hestroffer et al. 2019), which enables processes not observed in the laboratory. For example, contrary to expectations, the surface layers of Bennu (Barnouin et al. 2019, Lauretta et al. 2022, Walsh et al. 2022) and Ryugu (Sugita et al. 2019) were found to have very little to no cohesive strength. In this context, numerical simulations play an important role, as they enable modeling in physical environments that are not achievable in the laboratory. However, the realism of numerical simulations is limited by the models implemented and simulation parameters, which are typically calibrated upon laboratory-based results. The full-scale DART experiment provides an important ground-truth reference that may enable more realistic modeling of the physics of the impact as well as more informed interpretation of results in the view of generalizing and predicting dynamical behavior for other scenarios beyond DART.

## 4    CONCLUSIONS

It is too much to expect that a single kinetic impact experiment will uniquely constrain and accurately predict the outcome of a future kinetic impact on a different asteroid, in a different geometry, with a different spacecraft. Extrapolating the knowledge gained from DART to a new, and in detail unpredictable, situation will require reliance on physical models and computational techniques, for which the DART experiment forms an anchor point.  In this paper we have demonstrated how the measurement of the period change of Dimorphos's orbit caused by the DART impact directly constrains a direction-specific $\beta_p$, which in turn constrains the magnitude and direction of the ejecta momentum by straightforward geometry. We have shown how, using hydrodynamic simulations, this geometrical knowledge can be turned into constraints on material physical properties such as density, porosity, and cohesive strength. The computational expense of creating a complete library of validated simulations that fully spans the relevant parameter space still presents a challenge, although innovative techniques, including machine-learning interpretation of first-principles simulations, may speed progress. Developing this computational capability is key, though, as it will form the basis of our ability to predictively constrain the behavior of a yet-unknown asteroid surface.

In a hypothetical future where a kinetic-impact deflection of a genuinely dangerous asteroid becomes a necessity, gathering as much information as possible about the physical properties of the asteroid in advance of any attempt to deflect it will be paramount. We have discussed the level of information that might be available from Earth-based observations alone or supplemented by spacecraft flying by or rendezvousing with the target. Geophysical and compositional information will allow a narrowing of the parameter space and a sharpening of computational predictions for how the surface will respond. We also demonstrated that a direction-specific $\beta_u$ associated with $\hat{\mathbf{u}}$, the optimal direction for deflecting the asteroid, is a valuable figure-of-merit for mission design and execution. Given the likely range of behaviors for the ejecta response, strategically targeting a particular location on the asteroid to maximize the momentum transfer in the $\hat{\mathbf{u}}$ direction is both possible and potentially critical, as an impact in the wrong location could squander the effort or even make matters worse.

Several physical effects deserve further study, as they could prove to be important in specific situations; these include the various influences of rotation, and the escape of slow ejecta from binary systems. Notwithstanding the open avenues for future research, it will be wise to keep in mind that a real planetary defense situation may have unique aspects that could limit the range of possible action, for example, a heliocentric orbit that hinders easy access by spacecraft or opportunities for observation. The time and resources needed to develop and launch reconnaissance and impactor spacecraft may also limit what is possible. Exactly what degree of preparation should be maintained for planetary defense, whether in the form of ready-to-build designs on the shelf, ready-to-fly hardware in storage, or operating spacecraft in parking orbits, is not purely a scientific or engineering question, but a policy issue for space agencies and governments worldwide.

## ACKNOWLEDGEMENTS

The authors are grateful to Gareth Collins, Martin Jutzi, Emma Rainey, and Derek Richardson for critical readings of the manuscript at various stages. This work was supported in part by the DART mission, NASA Contract no. 80MSFC20D0004 to JHU/APL. Some of this work was carried out at the Jet Propulsion Laboratory, California Institute of Technology, under a contract with the National Aeronautics and Space Administration (80NM0018D0004). S.C. acknowledges funding through the Crosby Fellowship of the Department of Earth, Atmospheric and Planetary Sciences, Massachusetts Institute of Technology. E.D., S.I., A.L., M.P., A.R., and P. T. are grateful to the Italian Space Agency (ASI) for financial support through Agreement No. 2019-31-HH.0 in the context of ASI's LICIACube mission, and Agreement No. 2022-8-HH.0 for ESA's Hera mission. R. L. appreciates the funding from the European Union's Horizon 2020 research and innovation programme, grant agreement No. 870377 (project NEO-MAPP). P.M. acknowledges funding support from the French space agency CNES, ESA and the European Union's Horizon 2020 research and innovation program under grant agreement No. 870377 (project NEO-MAPP). R.M. acknowledges support from a NASA Space Technology Graduate Research Opportunities (NSTGRO) Award (80NSSC22K1173). R. N. acknowledges support from NASA/FINESST (NNH20ZDA001N). J. O. and I. H. were supported by the Spanish State Research Agency (AEI) Project No. MDM-2017-0737 Unidad de Excelencia "María de Maeztu" – Centro de Astrobiología (CSIC-INTA). They are also grateful for all logistical support provided by Instituto Nacional de Técnica Aeroespacial (INTA). S. R. S. acknowledges support from the DART Participating Scientist Program, grant no. 80NSSC22K0318. J. M. T. R. was funded by FEDER/Ministerio de Ciencia e Innovación - Agencia Estatal de Investigación of Spain (Grant No. PGC2018-097374-B-I00).